\documentclass[twoside]{dis07}
\usepackage[latin1]{inputenc}
\usepackage[dvips]{graphicx,epsfig,color}
\usepackage{wrapfig,rotating}
\usepackage{amssymb,amsmath,array}

\begin{document}
\title{Deeply Virtual Compton Scattering at HERA II}

\author{Laurent Schoeffel
%
\vspace{.3cm}\\
%
CEA Saclay, DAPNIA-SPP, 91191 Gif-sur-Yvette Cedex, France
}

\maketitle

\begin{abstract}
A new measurement is presented of elastic deeply virtual Compton scattering (DVCS) based
on data taken by the H1 detector during the complete HERA II period. 
The data are well described by QCD based calculations.
For the first time, a beam charge asymmetry  is obtained in a colliding mode, using data
recorded in $e^-p$ and $e^+p$. A significant non zero value is measured, related to
the interference of QCD and QED processes, namely the DVCS and Bethe-Heitler reactions.

\end{abstract}

\section{Introduction}

The DVCS reaction, $\gamma^* p \rightarrow \gamma p$,  can be interpreted as the elastic scattering of the
virtual photon off the proton via a colourless exchange, producing a real photon in the final state. 
It has a clear experimental signature identical to that of the purely 
electromagnetic Bethe-Heitler (BH) process. 
Since these two processes have an identical final state, it follows that they can interfere.
The squared photon production amplitude is then given by \cite{phibel}
\begin{equation} \label {eqn:tau}
\left| \tau \right|^2 
= \left| \tau_{{\scriptscriptstyle BH}} \right|^2 + 
\left| \tau_{{\scriptscriptstyle DVCS}} \right|^2 + \underbrace{
\tau_{{\scriptscriptstyle DVCS}} \, \tau_{{\scriptscriptstyle BH}}^* 
+ \tau_{{\scriptscriptstyle DVCS}}^* \, \tau_{{\scriptscriptstyle BH}}}_I,
\end{equation}
where $I$ denotes the interference term.
For an unpolarised proton target and lepton beam, the interference term can be written
quite generally as function of the azimuthal angle
$\phi$, the angle between the plane containing the incoming and outgoing leptons 
and the plane defined by the virtual and real photon \cite{phibel}
\begin{eqnarray} \label {I}
I \propto 
-C \, 
[  \cos \phi \, \mathrm{Re} \widetilde {\cal M}^{1,1} \, +
\cos 2\phi \, \mathrm{Re} \widetilde {\cal M}^{0,1} \, +
\cos 3\phi \, \mathrm{Re} \widetilde {\cal M}^{-1,1}
]
\label{int}
\end{eqnarray}  
where $C = \pm 1$ is the lepton beam charge and ${\cal M}^{i,j}$ are related to helicity amplitudes  \cite{phibel}.
Hence, cross section measurements which are integrated over $\phi$ are not sensitive 
to the interference term but
 the measurement of a cross section asymmetry with respect to the beam
charge
is a way to single out the interference term.

A general interest of the DVCS reaction lies in the the mass
difference (skewing) between the incoming virtual photon and the outgoing real
photon. This skewing can
 be interpreted in the context of generalised
parton distributions (GPDs) \cite{freund2} or in the dipole model framework \cite{lolo}.
In the following, new DVCS cross section measurements are presented and compared to QCD based models,
extending previous analyses \cite{dvcsh1,dvcszeus}.
For the first time, a beam charge asymmetry  is obtained in a colliding mode, using data
recorded in $e^-p$ and $e^+p$, during the HERA II data taking.

\section{Data analysis and results}

The measurements of the DVCS cross section are based on the data collected by the H1 detector 
during the years 2004 till 2007, 
with HERA running with positrons/electrons colliding protons of energy 27.6 and 920 GeV. It 
 corresponds to an integrated luminosity of 145 pb$^{-1}$ for each beam charge. 
To enhance the ratio of DVCS events w.r.t. BH ones, the photon is required to be detected
in the forward or central region of the H1 detector,
with a transverse momentum $P_T > 2$ GeV, while the scattered lepton is detected in the backward region, with an energy
$E > 15$ GeV. To ensure the elastic selection and reduce the proton dissociation background, the absence of activity
in the forward detectors is required \cite{dvcsh1}.   
To extract the DVCS cross section, the BH and inelastic DVCS backgrounds are subtracted bin by bin
and the data are corrected for trigger efficiency, detector acceptance and initial state photon radiation.

\begin{figure}[htbp] 
  \begin{center}
    \includegraphics[width=8cm]{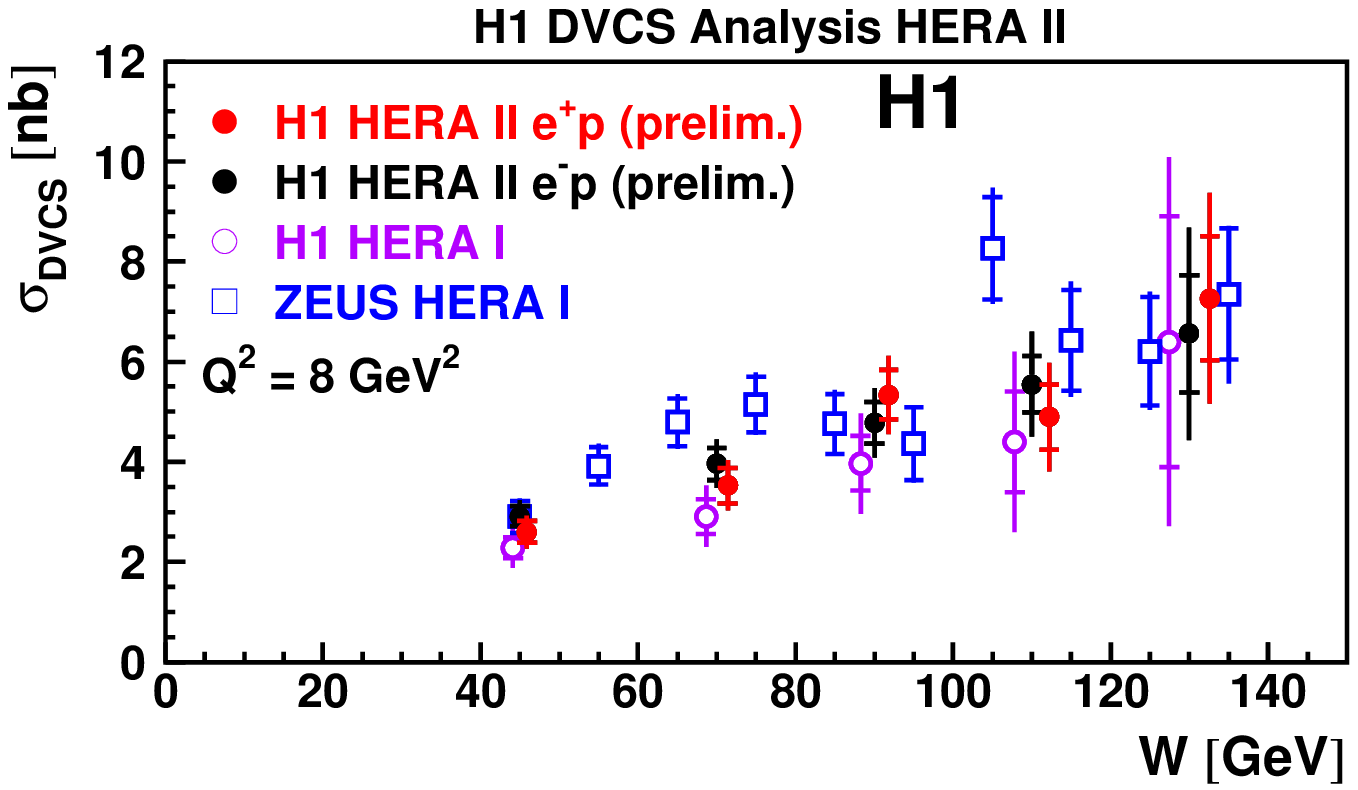}
    \includegraphics[width=5cm]{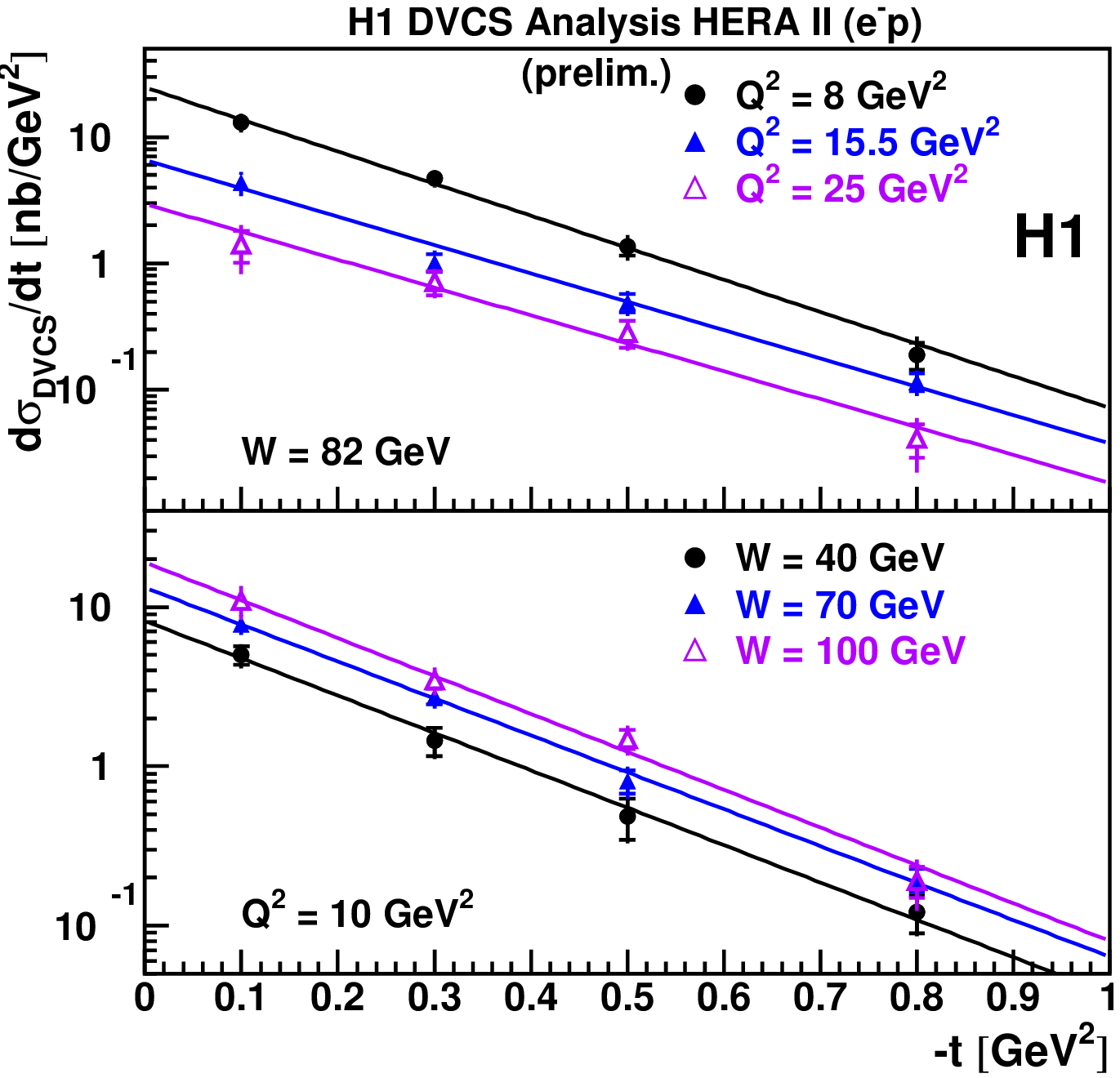}
  \end{center}
  \caption{DVCS cross section for positrons/electrons samples as a function of
$W$ (left) and differential in $t$, for 3 values of $Q^2$  and $W$ (right). 
 The results of a fit of the form $e^{-b|t|}$ are also displayed.
}
\label{fig1}  
\end{figure}

Results are presented in figure \ref{fig1}. In figure \ref{fig1} (left), we notice the reasonnable agreement of all analysis samples
for the dependence of the DVCS cross section as a function of 
the centre-of-mass energy of the $\gamma^*p$ system, $W$.
Using the electron sample only, the typical statistical and systematical errors on cross section are 10 \% and 15 \% respectively. Then, we can work with this electron sample for all analyses based on cross section.

In figure \ref{fig1} (right), we observe  the good description
of $d\sigma_{DVCS}/dt$ by a fit of the form $e^{-b|t|}$. Hence, an extraction of the $t$-slope parameter $b$ is accessible
for 3 values of the exchanged photon
virtuality, $Q^2$ and $W$, extending the previous determinations \cite{dvcsh1}. 
The global value of $b$ is found to be $5.45 \pm 0.19 \pm 0.34$~GeV$^{-2}$ at
 $W=82$~GeV and $Q^2=8$~GeV$^2$.
No dependence in $W$ is observed for $b$ and a significant $Q^2$ dependence can be extracted  using also previous measurements at lower $Q^2$ \cite{dvcsh1}.
We obtain :
$
b(Q^2)=A \left( 1-B \log(Q^2/2) \right)
$,
with 
$A=6.98 \pm 0.54$ GeV$^{-2}$ and $B=0.12 \pm 0.03$. 

\section{QCD Interpretations}

The  DVCS  cross section integrated over the momentum transfer can be written as

\begin{equation}
 \sigma_{DVCS} (Q^2,W)
  \equiv   \frac{\left[ \,Im{{A}}\,(\gamma^*p \to \gamma 
   p)_{t=0}(Q^2,W)\right]^2 (1+\rho^2)}{16\pi\,b(Q^2,W)}
\label{test1}
\end{equation}

\noindent
where $\rho^2$ is a small
correction due to the real part of the amplitude and~\cite{freund2}.
In the GPD formalism, 
the amplitude $Im {A}(\gamma^*p \to \gamma p)_{t=0}$ is directly proportional to the GPDs.

\begin{figure}[htbp]
  \begin{center}
    \includegraphics[width=7cm]{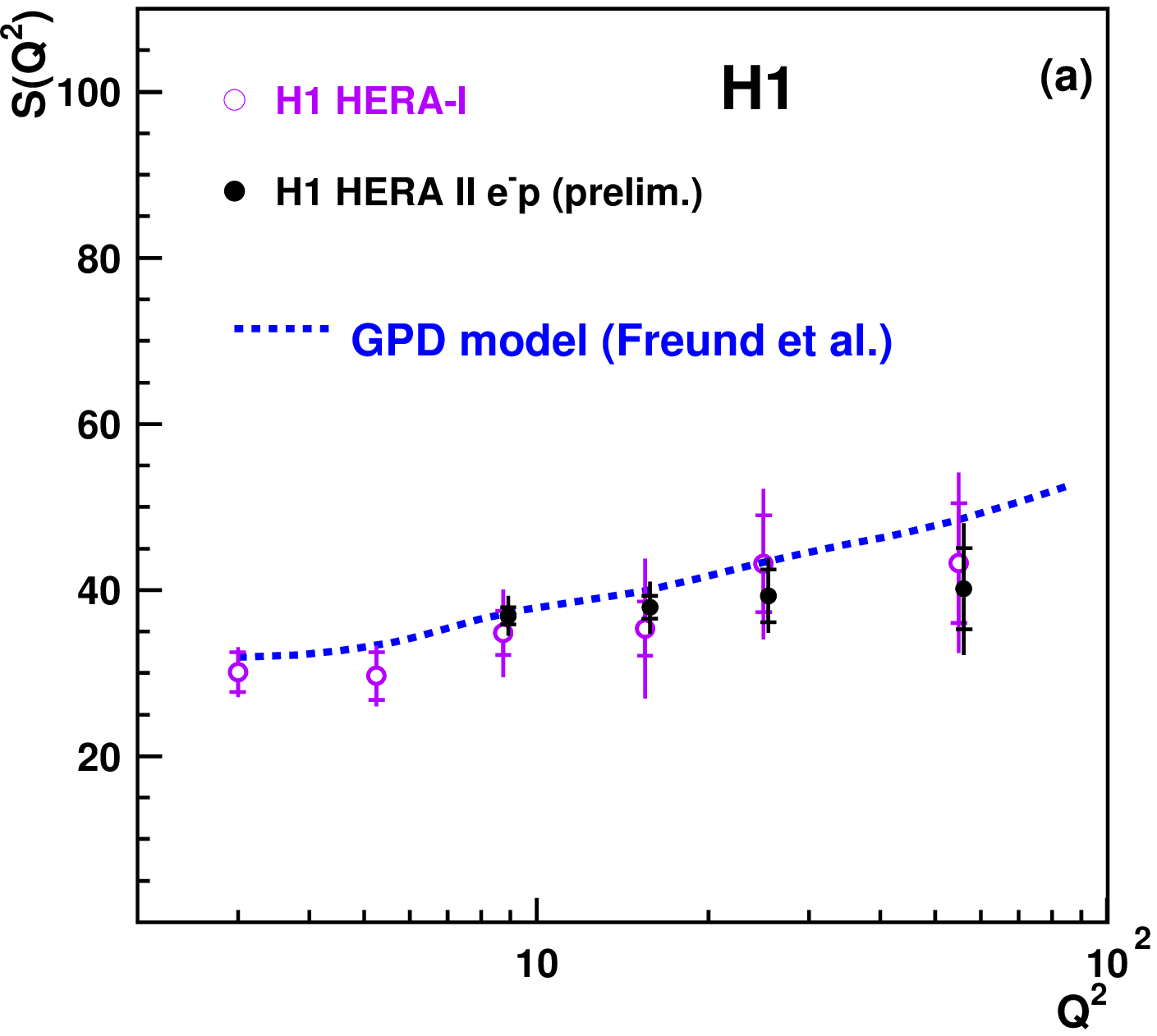}
    \includegraphics[width=5cm]{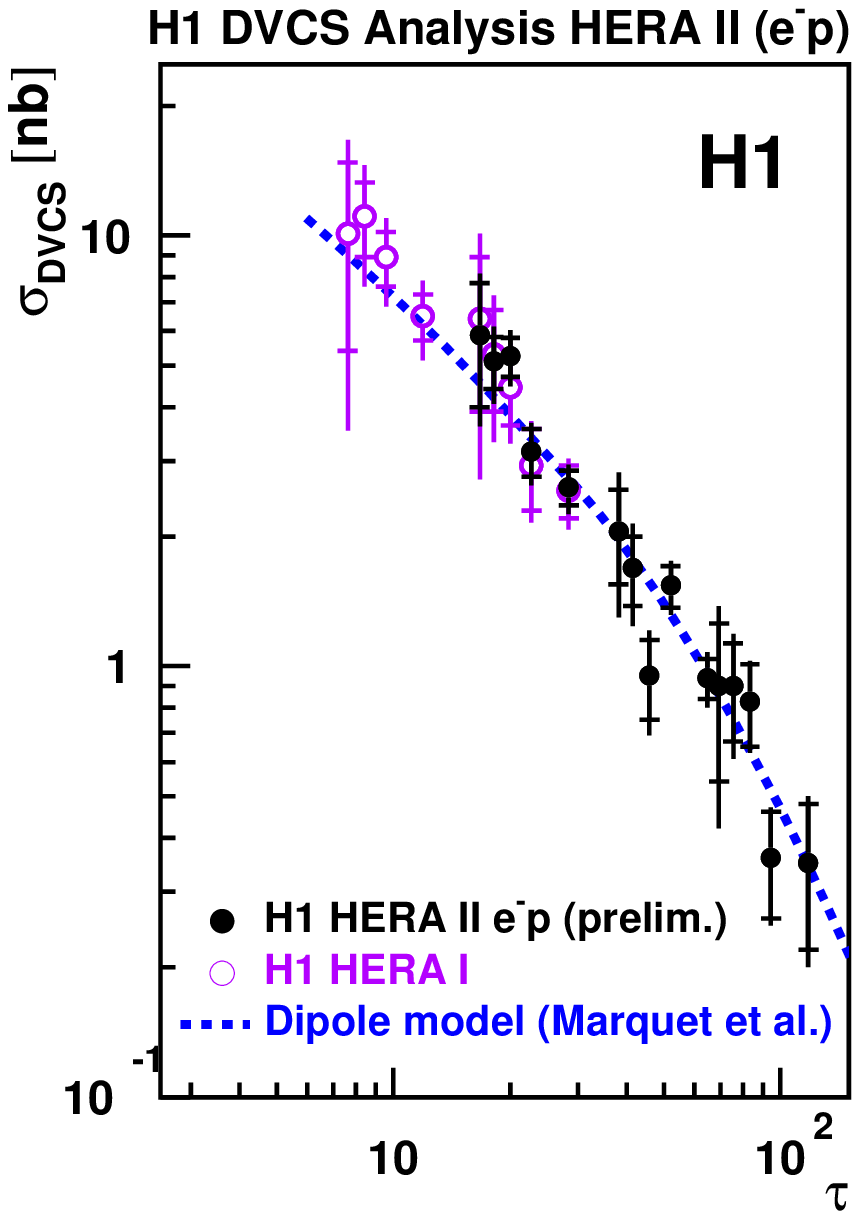}
  \end{center}
  \caption{Observable $S = \sqrt{ \frac{{\sigma_{DVCS} \ Q^4 \ b(Q^2)}} {{(1+\rho^2)}}}$ 
with the prediction for the GPD model \cite{freund2} (left).
DVCS cross section measurements as a function of $\tau={Q^2}/{Q_s^2(x)}$
with the prediction for the dipole model \cite{lolo} (right).
}
\label{fig2}
\end{figure}

We define
$
S = \sqrt{ \frac{{\sigma_{DVCS} \ Q^4 \ b(Q^2)}} {{(1+\rho^2)}}} 
$, which is proportional to $|Im{{A}}\,(\gamma^*p \to \gamma p)_{t=0}(Q^2,W)|$
and  therefore directly contains information on the $Q^2$ evolution for the GPDs.
The result is shown in figure \ref{fig2} (left)  an compared to a GPD model~\cite{freund2}.
A reasonable description of the weak $Q^2$ dependence,
compatible with a logarithmic behaviour, is observed for $S$.

The DVCS cross section can also be interpreted within the dipole approach~\cite{lolo}.
It expresses the 
scattering of the virtual photon off the proton through its fluctuation into a color singlet 
$q\bar q$ pair (or dipole) of a transverse size $r\!\sim\!1/Q$. 
In the  dipole approach 
the DVCS
cross section is expected to verify the genuine property of 
geometric scaling~\cite{lolo}. This means that the cross section does not depend
 on both $x \simeq Q^2/W^2$ and $Q^2$ but obeys a scaling in a single variable $\tau=Q^2/Q^2_s$, where 
 $Q_s(x)=Q_0 ({{x}/{x_0}})^{-\alpha/2}$. Using parameters $Q_0= 1$ GeV, $\alpha=0.25$ and $x_0=2.7\ 10^{-5}$~\cite{iim2},
we can determine the variable $\tau$. The DVCS cross section is presented 
as a function of $\tau$ in figure \ref{fig2} (right), which indicates that 
the geometric scaling property is verified within the present errors.

\section{Beam charge asymmetry }
\begin{figure}[htbp] 
  \begin{center}
    \includegraphics[width=10cm]{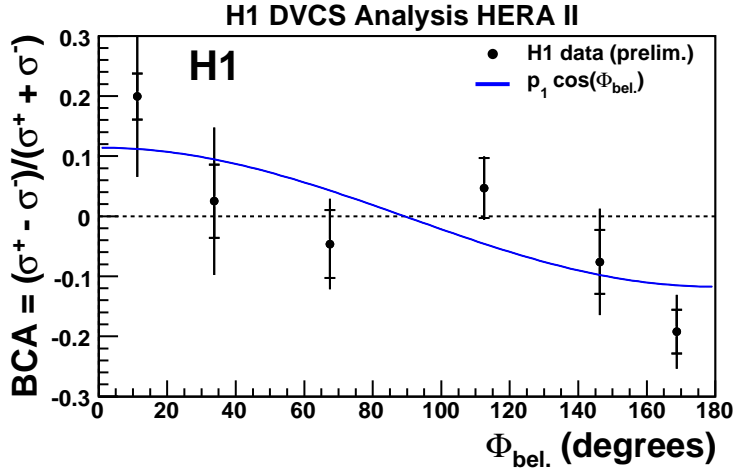}
  \end{center}
  \caption{Beam charge asymmetry as a function of $\phi$ \cite{phibel}.
}
\label{fig3}  
\end{figure}

The determination of a cross section asymmetry with respect to the beam
charge is realised by measuring the ratio
$(d\sigma^+ -d\sigma^-)/ (d\sigma^+ + d\sigma^-)$ as a function of $\phi$.
Note that $\phi$ is not defined when $|t|<|t|_{min}=x^2m_p^2/(1-x)$  \cite{phibel}. However, 
the experimental resolution in $t$ is larger than the kinematical limit $|t|_{min}$. Therefore
we can not define $\phi$ when $|t|<0.05$ GeV$^2$ and the BCA is measured only
for  $|t| \ge 0.05$ GeV$^2$.
In the  expression of the BCA, $d\sigma^+$ and $d\sigma^-$ correspond to the DVCS cross section
measured in positron and electron samples, over a bin $d\phi$
\footnote{
The azimuthal angle $\phi$ 
is defined according to the convention of \cite{phibel}.}.
Results are presented in figure \ref{fig3} with  a fit in $\cos \phi$,
which is expected to be the first dependence in $\phi$ following equation (\ref{int}).
After applying a deconvolution method to account for the  resolution on $\phi$,
the coefficient of the $\cos \phi$ dependence is found to be $p_1 = 0.17 \pm 0.03 (stat.) \pm 0.05 (sys.)$.

%
%
%
 

\section{Conclusion}

The DVCS cross section 
has been measured over a large kinematic domain using the complete HERA II data,
extending previous analyses \cite{dvcsh1, dvcszeus}.
For the first time, a beam charge asymmetry  is obtained in a colliding mode, using data
recorded in $e^-p$ and $e^+p$. A significant non zero value is measured for  $|t| \ge 0.05$ GeV$^2$, 
which is related to
the interference of DVCS and BH processes. 

\begin{footnotesize}



%

\end{footnotesize}


\end{document}